\documentclass[sigconf]{acmart}

\AtBeginDocument{%
  }

\setcopyright{acmlicensed}
\copyrightyear{2025}
\acmYear{2025}

\acmConference[W4A '25]{Proceedings of 22nd International Web for
All Conference (W4A ’25). Sydney, Australia, 5 pages, to appear}{April 28--29,
  2025}{Sydney, Australia}
\acmISBN{}

\begin{document}


\title[Deaf Multimedia Authoring Tax]{Barriers to Employment: The Deaf Multimedia Authoring Tax}


\author{Christian Vogler}
\email{christian.vogler@gallaudet.edu}
\orcid{0000-0003-2590-6880}
\affiliation{%
  \institution{Gallaudet University}
  \city{Washington}
  \state{D.C.}
  \country{USA}
}

\author{Abraham Glasser}
\email{abraham.glasser@gallaudet.edu}
\orcid{0000-0003-1763-4352}
\affiliation{%
  \institution{Gallaudet University}
  \city{Washington}
  \state{D.C.}
  \country{USA}
}

\author{Raja Kushalnagar}
\email{raja.kushalnagar@gallaudet.edu}
\orcid{0000-0002-0493-413X}
\affiliation{%
  \institution{Gallaudet University}
  \city{Washington}
  \state{D.C.}
  \country{USA}
}

\author{Matthew Seita}
\email{matthew.seita@gallaudet.edu}
\orcid{0000-0001-7991-2704}
\affiliation{%
  \institution{Gallaudet University}
  \city{Washington}
  \state{D.C.}
  \country{USA}
}

\author{Mariana Arroyo Chavez}
\email{marroyoch22@gmail.com}
\orcid{0009-0004-6114-4355}
\affiliation{%
  \institution{Gallaudet University}
  \city{Washington}
  \state{D.C.}
  \country{USA}
}

\author{Keith Delk}
\email{kdelk22@gmail.com}
\orcid{0009-0004-5265-6586}
\affiliation{%
  \institution{Gallaudet University}
  \city{Washington}
  \state{D.C.}
  \country{USA}
}

\author{Paige Devries}
\email{paige.devries@gallaudet.edu}
\orcid{0009-0009-9815-7121}
\affiliation{%
  \institution{Gallaudet University}
  \city{Washington}
  \state{D.C.}
  \country{USA}
}

\author{Molly Feanny}
\email{mkfeanny12@gmail.com}
\orcid{0009-0005-9889-6355}
\affiliation{%
  \institution{Gallaudet University}
  \city{Washington}
  \state{D.C.}
  \country{USA}
}

\author{Bernard Thompson}
\email{bernard.thompson@gallaudet.edu}
\orcid{0009-0008-1611-4287}
\affiliation{%
  \institution{Gallaudet University}
  \city{Washington}
  \state{D.C.}
  \country{USA}
}

\author{James Waller}
\email{james.waller@gallaudet.edu}
\orcid{0000-0002-8562-8336}
\affiliation{%
  \institution{Gallaudet University}
  \city{Washington}
  \state{D.C.}
  \country{USA}
}

\renewcommand{\shortauthors}{Vogler et al.}

\begin{abstract}
  This paper describes the challenges that deaf and hard of hearing people face with creating accessible multimedia content, such as portfolios, instructional videos and video presentations. Unlike content consumption, the process of content creation itself remains highly inaccessible, creating barriers to employment in all stages of recruiting, hiring, and  carrying out assigned job duties. Overcoming these barriers incurs a "deaf content creation tax" that translates into requiring significant additional time and resources to produce content equivalent to what a non-disabled person would produce. We highlight this process and associated challenges through real-world examples experienced by the authors, and provide guidance and recommendations for addressing them.
\end{abstract}



\received{8 March 2025}
\received[revised]{25 March 2025}
\received[accepted]{25 March 2025}

\maketitle

\section{Introduction}

In this communication paper, we describe the challenges around multimedia authoring, as experienced firsthand by a mixed deaf, hard of hearing and hearing (DHH/H) team. Most discussions around accessibility barriers for DHH individuals in the workplace, as well as broader society, have focused on communication barriers, and the \textbf{consumption of content.} These are typically accommodated through sign language interpreters, captions/subtitles, and transcripts. However, there is a much underappreciated problem: \textbf{the process of creating content} itself is not accessible. This holds doubly for content that is required to be fully compliant with guidelines in the United States (Section 508), in the European Union, Canada and Australia (EN 301 549), and internationally (Web Content Accessibility Guidelines - WCAG), as it forces DHH workers to deal with modalities that are inaccessible to them, such as voiceover audio descriptions for the blind. The premise of this paper is that DHH people pay a huge ``deaf multimedia authoring tax'' in terms of expended time and resources, which their non-disabled peers and workplace colleagues simply do not face.

Content creation is a cornerstone of the modern workplace, spanning the lifecycle from job hunting to hiring to carrying out assigned duties. A person looking for a job may have to create a professional portfolio with multimedia content. They may have to create multimedia segments as part of the hiring and interview process. On the job, they may have to create accessible presentations, instructional videos, social media posts, demonstration videos, internal communications and more. A DHH person faces barriers in each area, due to inaccessible content creation processes. They may not be able to easily create a compelling multimedia portfolio and get passed over in recruiting. They may not be able to satisfy the demands of the hiring process and get rejected. They may not be able to carry out the content creation duties that are mission-critical to the employer and get reassigned, demoted, or fired. 

\subsection{Our Team with Lived Experience}
In the remainder of this paper, we describe through examples how our team of DHH/H individuals struggled with meeting accessibility requirements and mainstream expectations for professional video content. Our team consists of nine DHH people who all use American Sign Language (ASL) as the primary mode of communication. While the majority also uses spoken English, they all exhibit ``deaf accents'' that impede their ability to produce intelligible voiceovers for multimedia content~\cite{hudgins1934comparative,maassen1984effect,fok2018towards}. The team also includes one hearing person who has training as a sign language interpreter and is fluent in both ASL and spoken English. All team members also have excellent written English skills. One aspect that we wish to highlight in particular is that neither having sign language interpreters on the team, nor having excellent written English skills, lessen the challenges of accessible content creation. The ``deaf multimedia authoring tax'' persisted regardless.

\subsection{Related Work on DHH Content Creation}
Some past work has examined the experiences of DHH people for content creation on social media, such as creating captions~\cite{li2022exploration}. Another paper has looked at content created by and for deaf people~\cite{tang2023community} and suggested that creating accessible content should be community driven. However, while this approach may work for social media, it would face significant challenges with integrating it into the workplace, where individual contributions to the team dominate. Another paper found that TikTok engagement was closely correlated to whether the deaf-created video had voiceover, and also highlighted the significant challenges that DHH creators face with providing such voiceover~\cite{cao2024voices}. Another school of thought that has seen some debate in the DHH community is to let sign language stand on their own, by not providing captions, let alone voiceover, at all~\cite{hayden2023cognitive,limpingchickenQuestionShould}. One reason that has factored into these debates is the amount of effort that it takes for DHH content creators to add these to their videos. However, this debate has taken place in the context of social media and blogging platforms. In the workplace, in contrast, creators do not have a choice: the content they produce must both conform to mainstream expectations, with audio as one of the media, and also be fully accessible.

\section{Creating Videos for a Hearing-Dominated World}

In this section we are providing several examples related to video content production that have created difficult challenges for our DHH team. DHH people generally face barriers with producing videos, irrespective of the mode of communication, because of the above-mentioned deaf accents, as well as limited abilities to listen and perform quality control. However, these barriers are magnified for sign language users, due to the fact that ASL and spoken English are different languages, and due to the fact that ASL does not have a universally used written form.

A typical workflow for video production originating with a DHH content creator who uses ASL is:
\begin{enumerate}
    \item Create an initial English script for the content
    \item Translate the script from English into ASL, taking notes on how to sign specific concepts
    \item Create a draft ASL video with the prompts for the final ASL translation
    \item Use the draft video to film the final ASL content
    \item Translate the ASL back into an English transcript
    \item Align the timing of the English with the ASL in an SRT file
    \item Create English voiceover in one of two ways:
    \begin{enumerate}
        \item Engage a hearing person to speak the timed English
        \item Use text-to-speech (TTS) software to speak the timed English
    \end{enumerate}
    \item Mix the voiceover audio into the video
    \item Create an audio description script
    \item Record the audio descriptions (via a human or via TTS)
    \item Mix the audio descriptions into the video, taking care not to overlap with the voiceover
\end{enumerate}

As the steps above show, the workflow becomes much more complex compared to an English-originated video. Originating a video in ASL, then adding captions and English voiceover requires multiple steps, including translation, re-recording videos, and synchronization of the voiceover with both captions and the signing. Translation, in particular, is time-consuming --- creators must first sign naturally, then back-translate to English for captions and voiceovers. But signing naturally itself is difficult to achieve if the creator is to make use of any notes written in English. The English writing affects the production of signs, due to the effect of language contact between ASL and English~\cite{lucas1990asl}. In order to eliminate this effect, creators have to start with a draft translation video based on their English notes and then re-record the draft video to generate fully natural signing, as shown in steps (1)--(4).

Once the English to ASL translations are done, more work awaits in the form of back-translating the ASL to English, and then creating captions with timing aligned to the ASL, as shown in Steps (5)--(6). The latter currently cannot be automated, unlike with aligning English voiceover with captions, because there is no production-ready software that can translate between ASL and English. 

\begin{figure}[h]
  \centering
  \includegraphics[width=0.8\columnwidth]{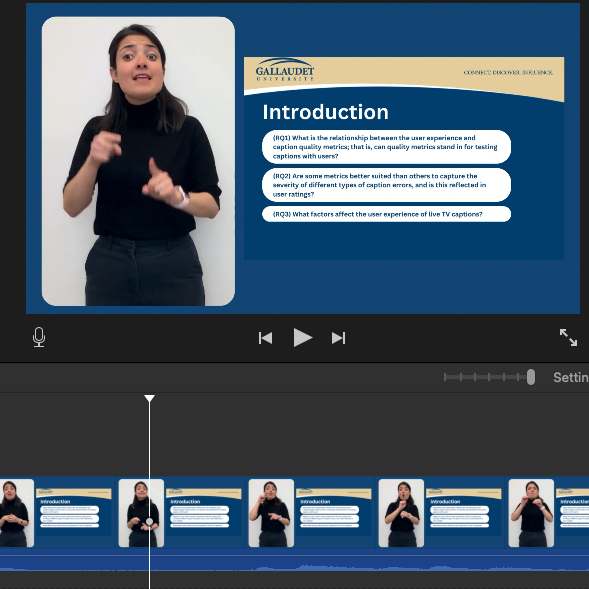}
  \caption{Screenshot of video editor showing a person signing actively, but the audio track of the voiceover shows silence at that moment.}
  \label{fig:silence}
  \Description{A video editor with a timeline. A young woman is in the middle of signing something. The timeline visualizes the audio, which is completely silent at the marked point in time.}
\end{figure}

The next step, generating the voiceover in steps (7)(a)/(b) and (8), faces unique challenges because the pacing between ASL and English is different. Much like with spoken language translation, there are concepts that can be expressed concisely in ASL, but require a long string of English words to express, and vice versa. This makes aligning the voiceover with the timing of the captions and the signing extremely difficult to accomplish. In our team's experience, one of two things almost always happen: First, the spoken English voiceover is faster than the ASL signing, and there are awkward slowdowns or pauses in the speaking, while the signing continues. This effect can be seen in a video on caption quality, created by the authors for the ACM CHI 2024 conference~\cite{chavez2024video}; for instance, at timecode 0:10, there is an awkward silence to allow the ASL to catch up, as shown in Figure~\ref{fig:silence}.

Second, and conversely, the signing can be so fast that the voiceover is unable to fit in the allotted time slots. In this situation, DHH creators are faced with an unenviable choice. They can either speed up the English voiceover recording to make it fit the timing, which also makes it more difficult to understand. Or they can extend the time the captions stay on screen, which results in the ASL and the English no longer being in sync. The impact of speeding up audio on intelligibility can be observed in another video created for CHI 2024, which adjusted the speaking rate of a TTS system to 1.35x speed at timecode 0:16~\cite{chavez2024preview}, while also lengthening the duration of some captions in the same video to give the English voiceover more time.

It is further important to note that caption delays relative to the audio and video create new accessibility problems.~\cite{burnham1998captions}. DHH people who listen to audio and watch captions at the same time are especially sensitive to timing mismatches~\cite{armstrong2013development}. In sum, the language translation component between ASL and English creates difficult-to-tackle barriers for DHH content creators to make their videos fully accessible. The problems with timing and synchronization persist irrespective of whether TTS is used to generate the voiceover, or whether a hearing interpreter is employed. Our DHH/H team has tried both approaches, and encountered this problem in both.

Finally, steps (9)-(11) are a critical component of making videos accessible to blind people, and also required for compliance with WCAG, Section 508 and EN 301 549. From a DHH perspective, these steps face timing and synchronization problems analogous to steps (7)--(8). However, they are caused by something entirely different --- a DHH creator is not able to plan ahead of time where the gaps in the voiceover are to allow filling in audio descriptions. These gaps are dependent on the timing of the voiceover, rather than the ASL signing, and thus cannot be accounted for until step (8) is complete. A DHH creator also cannot reliably estimate how long it will take to speak the audio description from a script. As a result, even careful planning and the best of intentions for making a video accessible to blind people can be derailed. This effect can be seen in another video presented at ACM CHI 2024~\cite{tran2024alexa}. The team had planned to include audio descriptions, and completed all recordings with our hearing team member for both the voiceover and the audio descriptions. But in the end, there was insufficient available timing to include the carefully planned audio descriptions. The team was forced to make the difficult decision to drop audio descriptions from the video. Fixing the problem would have required extensive editing of the video, or re-recording significant portions of the audio. In either case, the team would have run out of time and missed its deadlines.

\subsection{The Deaf Multimedia Authoring Tax Quantified}

We have established the challenges involved with producing videos as DHH content creators. In the following, we provide an estimated breakdown of the additional resources that were required as a result. Going back to the example of the full caption quality video presented at CHI 2024~\cite{chavez2024video}, our project notes and communication archives allow us quantify the time that it took for a team of four people to carry out each extra step involved in the production. All this was for a 15-minute video.

\begin{itemize}
    \item \textbf{6 person-hours} for Step (2) and (3) --- English to ASL translation and draft videos
    \item \textbf{2.5 person-hours} for Step (4) --- filming clean ASL
    \item \textbf{1 hour} for Step (5) --- back-translating the transcript
    \item \textbf{1 hour} for Step (6) --- creating and manually aligning captions
    \item \textbf{1 hour} for Step (7)(a) --- hearing human voiceover
    \item \textbf{1.5 hours} for Step (8) --- mixing  the voiceover and realigning with captions
\end{itemize}

This is a total of 13 hours spent above and beyond planning and editing the video. While a hearing person would have to spend some time filming a video with voiceover, equivalent to Step (4), and some time on generating captions, equivalent to Step (6), this is likely to take no more than 1--2 hours. In addition, because the audio is already present in a hearing-generated video, the caption creation step can be outsourced at a cost of U.S. \$1.25 to U.S. \$3 per minute, based on current rates. All in all, based on these figures, the ``deaf multimedia authoring tax'' is in excess of 10 hours, which is wildly disproportionate and puts DHH content creators at a significant disadvantage. There is an adage that people with disabilities have to work twice as hard to prove themselves~\cite{morina2017we}, which certainly would be true any time that multimedia content is required during job recruitment or job applications. And if multimedia content creation is part of an assigned task at a job, it makes DHH employees disproportionately more expensive at carrying out the task.

\subsection{Quality Control Challenges}

Aside from the extra time that DHH multimedia uthors need to spend on production, there are additional challenges with quality control. Some of these can be alleviated with a hearing team member who is able to listen to the audio, but our team has experienced situations where such help was limited or unavailable altogether.

In the previous section we mentioned that the voiceover in Step (7) can be created through either a hearing human person, or through TTS methods. The latter is the only option when no hearing person is available, or when a DHH team has to make last-minute edits.  TTS tools, even with state-of-the-art technologies that often require paid subscriptions, need more research and development to generate more natural-sounding English voiceovers. In addition, DHH users may not be able to evaluate effectively whether the generated speech is appropriate in delivery, emotion, and other prosodic information, due to insufficient usable hearing. Either they cannot hear it at all, or they may not have access to the acoustic features that convey it, especially with cochlear implants~\cite{everhardt2020meta}.

\begin{figure}[h]
  \centering
  \includegraphics[width=\columnwidth]{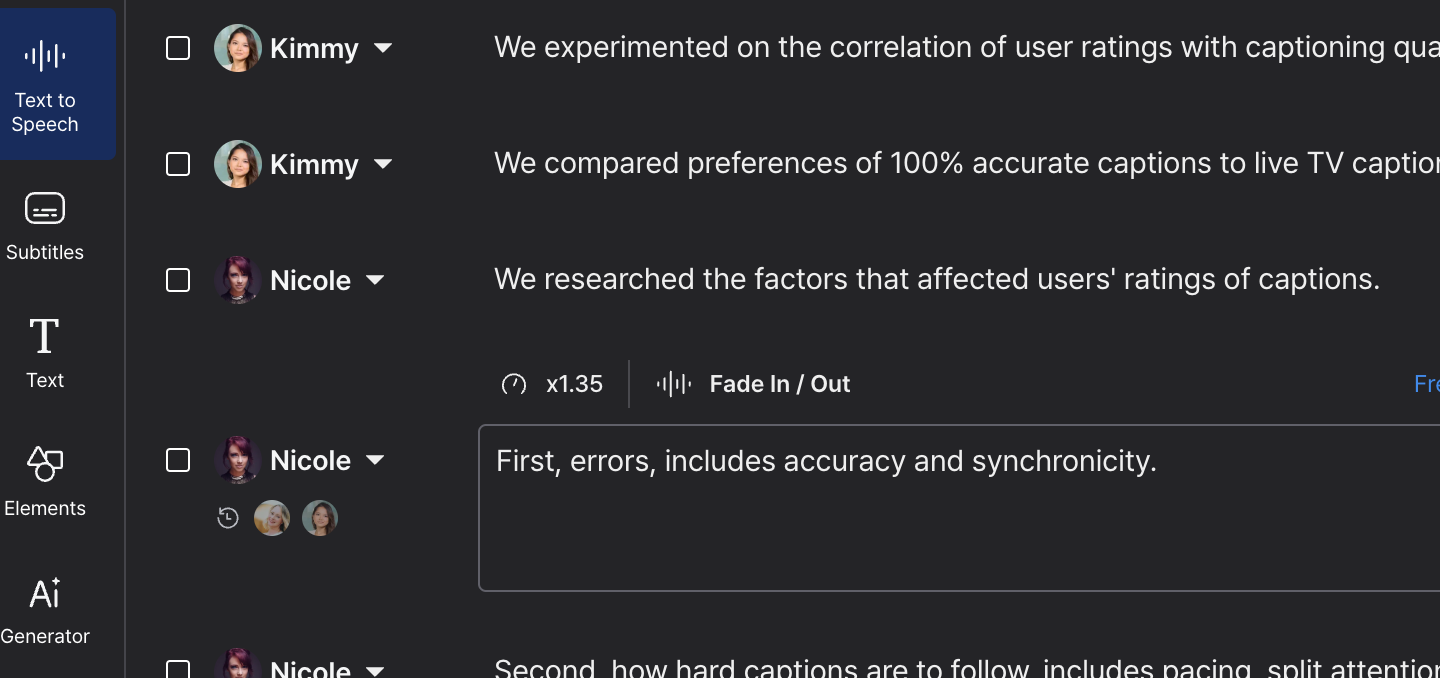}
  \caption{Screenshot of TTS editor showing how a text segment is sped up by a factor of 1.35x using an AI-generated voice. DHH creators struggle with assessing whether the result is intelligible.}
  \label{fig:lovo}
  \Description{A web window showing paragraphs stacked vertically and on a timeline. Each paragraph has an associated named AI voice. The highlighted paragraph features a time control setting it to 1.35x speed.}
\end{figure}

One concrete challenge with TTS delivery quality control is illustrated in Figure~\ref{fig:lovo} showing the addition of TTS voiceover. We previously mentioned that we had to speed up TTS by 1.35x to make the audio fit in the allotted space, even when allowing the caption timing to drift slightly. There is no way for a DHH person to judge reliably whether the 1.35x speed is intelligible to begin with. Although team members used their hearing aids and cochlear implants to listen, their judgment likely was affected by having prior knowledge of the exact words. 

Another quality control challenge has arisen in our DHH team with hearing collaborators. A co-author of this paper who is a Deaf signer had an incident where a hearing team member, not fluent in ASL, led the video production and made a video where the ASL portion was backward. While this would have been immediately obvious and nonsensical for an experienced ASL learner or a native/fluent signer, it still looks like ASL for naive signers and people who do not know ASL. To make sure that such mistakes are caught, it is crucial for a hearing person who is fluent in ASL to conduct quality control by watching the media and making sure the different components (ASL content, English audio voiceover and descriptions, and English captions) are aligned and understandable.

\section{Future Directions}

Given the importance of multimedia in the modern workplace, it essential to reduce the ``deaf multimedia authoring tax'' burden, in order to allow DHH people to stay competitive. Our experience as a DHH/H team has shown that having hearing team members alone is not sufficient. The inaccessibility of the content creation process is just too pronounced. However, there is potential for major advances if content creation toolchain developers started to focus on ensuring that their workflows are fully accessible to DHH people. 

Sign language-aware tools would be especially helpful. Sign language technologies, particularly sign language recognition and translation \cite{sl_interdisciplinary}, would be useful in streamlining support for a signed-language workflow, rather than a spoken-first approach. For instance, sign language translation could help with step (5), translating the ASL content into an English transcript. Sign language recognition and translation technology could also be used to address steps (6) and (8) to align the timing of the English with the ASL into a subtitle file, and mixing the voiceover audio into the video.

DHH people should also be able to customize AI-based TTS voices and manner of expression to their liking and have the ability to verify that the output is what it is supposed to be. They further should be able to employ text-to-speech models that reflect their own unique personalities, not dissimilar to how mainstream providers like Eleven Labs and Lovo.ai support voice cloning.

AI-based tools integrated into the creation platforms also could potentially be of major help with verification. A simple step would be built-in automatic speech recognition to verify the timing and content of the audio with respect to the caption/subtitle files that were created in step (6). More advanced tools could potentially detect speaker changes, provide speaker identification, and provide a description of the voice characteristics, delivery and manner of speaking, to ensure that it aligns with the characteristics and delivery of a person signing.

Beyond technical solutions, there also needs to be a recognition of how inaccessible workflows are. In general, employers and platforms should recognize that requiring voiceover in video applications and content places an undue burden on deaf signers and either offer alternative methods for submitting content, or offer the necessary support to generate high-quality voiceovers.

\section{Conclusion}

In this paper we have described the challenges that DHH people face with content creation in a hearing-centric voice-oriented world. It is no exaggeration to state that there is a ``deaf multimedia authoring tax,'' which in one example was calculated to require at least 10 extra hours of work. There are potential solutions that may be feasible today or in the near future by enhancing the technical capabilities of content creation tools. Until this problem is addressed, however, DHH people will be at significant competitive disadvantages in modern employment contexts. 

\begin{acks}
The contents of this paper were developed under a grant from the National Institute on Disability, Independent Living, and Rehabilitation Research (NIDILRR grant number 90REGE0027). NIDILRR is a Center within the Administration for Community Living (ACL), Department of Health and Human Services (HHS). The contents of this paper do not necessarily represent the policy of NIDILRR, ACL, HHS, and you should not assume endorsement by the Federal Government.  Additional support was provided by National Science Foundation grants \#2150429, \#2348221, and \#2447704.
\end{acks}

\bibliographystyle{ACM-Reference-Format}
\bibliography{bibliography}

\appendix

\end{document}